\documentclass[showkeys]{revtex4-2}
\usepackage{tabularray}
\usepackage{amsmath,graphicx,array,float,amsthm,amssymb}
\usepackage{tikz,color,fullpage,epsf,caption,subcaption,multirow}
\usepackage{amsmath,graphicx,verbatimbox,array,float}
\usepackage{indentfirst,orcidlink}
\usepackage{braket}
\usepackage{diagbox}
\usepackage{amssymb}
\usepackage{bbold}

\renewcommand{\eqref}[1]{Eq.~(\ref{#1})}
\newcommand{\tr}{{\mathrm{Tr}}}

\renewcommand{\eqref}[1]{Eq.~(\ref{#1})}

\begin{document}
\title{Quantum State Tomography of Photonic Qubits with Realistic Coherent Light Sources}
\author{Artur Czerwinski \orcidlink{0000-0003-0625-8339}}\email{aczerwin@umk.pl}
\affiliation{Institute of Physics, Faculty of Physics, Astronomy and Informatics, Nicolaus Copernicus University in Torun, ul. Grudziadzka 5, 87-100 Torun, Poland}

\begin{abstract}
Quantum state tomography (QST) is an essential technique for characterizing quantum states. However, practical implementations of QST are significantly challenged by factors such as shot noise, attenuation, and Raman scattering, especially when photonic qubits are transmitted through optical fibers alongside classical signals. In this paper, we present a numerical framework to simulate and evaluate the efficiency of QST under these realistic conditions. The results reveal how the efficiency of QST is influenced by the power of the classical signal. By analyzing the fidelity of reconstructed states, we provide insights into the limitations and potential improvements for QST in noisy environments.
\end{abstract}

\keywords{Quantum state tomography, shot noise, coherent states, Raman scattering, channel crosstalk}

\maketitle

\section{Introduction}

Quantum state tomography (QST) is a fundamental tool in quantum information science, providing a means to fully characterize the quantum state of a physical system \cite{Paris2004,Toninelli2019,Czerwinski2022,Gebhart12023}. It serves as a bridge between experimental measurements and the theoretical description of quantum systems, enabling us to verify and validate the behavior of quantum devices. Accurate knowledge of quantum states is essential for a wide range of applications, including quantum computing, quantum communication, and quantum cryptography \cite{Walther2005,Kok2007,Wei2022,Slaoui2023,Slaoui2024,Ali2024}. As these technologies advance, the ability to reliably reconstruct quantum states from experimental data becomes increasingly important.

However, the practical implementation of QST is challenged by various imperfections that arise during the transmission of quantum states. Although photonic states are preferred for their practical applications \cite{James2001,Altepeter2005,Horn2013,Temporao2024}, they encounter limitations when transmitted through optical fibers. As quantum states propagate through optical channels, they are subject to attenuation, which reduces the number of photons that can reach the detection system, and to various types of noise, such as shot noise and Raman scattering \cite{Eraerds2010,Hasinoff2014,Frohlich2015}. These factors can significantly distort the measured data, leading to inaccuracies in the reconstructed quantum states. Attenuation in optical fibers, for instance, causes an exponential decay in the signal strength with distance, while shot noise introduces random fluctuations in the photon count due to the inherent quantum nature of light. For this reason, developing and testing frameworks that account for various sources of measurement uncertainty remains an important area of ongoing research, see, for example, Refs. \cite{CzerwinskiQIC,Wang2022,CzerwinskiPRA,Pedram2024}.

The concept of this paper is directly related to wavelength-division multiplexing (WDM), which is one of the key technologies in modern optical communication systems. WDM allows multiple optical carrier signals to be transmitted simultaneously through a single fiber by utilizing different wavelengths for each signal \cite{Mukherjee2006}. This technique greatly increases the capacity and efficiency of optical networks, making it possible to transmit large volumes of data over long distances. However, when quantum signals are transmitted alongside classical signals in the same fiber, as in WDM systems, the interaction between these signals can introduce significant noise into the quantum channel \cite{Gobby2004,Tomita2010,Cao2019}. A particular concern is Raman scattering, a nonlinear optical process where photons from the classical signal scatter and generate noise in the quantum channel, potentially degrading the quality of the quantum signal. Raman scattering further complicates the measurement by introducing additional noise that can mask the true quantum signal. This motivates the study of Raman scattering's impact on QST, as understanding and mitigating this noise is crucial for the reliable reconstruction of quantum states in practical WDM-based quantum communication systems.

Another critical issue in WDM systems is channel crosstalk, where signals from adjacent wavelength channels interfere with each other due to imperfections in optical components, such as filters and multiplexers, or nonlinear effects within the fiber, including Raman scattering and four-wave mixing \cite{Zhou1996,Monroy2013}. This spectral crosstalk occurs when optical signals at different wavelengths interact, causing leakage of power between channels. In the context of quantum communication, crosstalk can introduce additional noise into the quantum channel, especially when classical signals are transmitted alongside quantum states \cite{Eraerds2010}. This noise complicates the accurate reconstruction of quantum states by obscuring the quantum signal, thus reducing the fidelity of QST. In this contribution, we investigate the impact of spectral crosstalk on QST, analyzing how interference from neighboring classical channels affects the fidelity of the reconstructed quantum states.

In this paper, we propose a numerical framework to simulate and evaluate the efficiency of QST under realistic conditions, taking into account shot noise, attenuation, and Raman scattering. Our framework allows for the generation of photon counts corresponding to a variety of noise scenarios, enabling a detailed analysis of the performance of QST in practical settings. We model the expected photon counts using the Born rule, incorporate the effects of fiber attenuation, and simulate the impact of Raman scattering on the measurements. This approach provides a comprehensive tool for assessing the robustness of QST in the face of these challenges.

The theoretical foundations of our framework are outlined in Section \ref{framework}, where we describe the mathematical modeling of realistic sources and the corresponding photon-counting schemes. In Section \ref{resultsanalysis}, we present the results of our numerical simulations, offering a detailed analysis of how attenuation and noise affect the fidelity of the reconstructed quantum states. By comparing the performance of QST under different conditions, we provide insights into the limitations of current techniques and suggest potential avenues for improving the accuracy of quantum state reconstruction in noisy environments. The paper is concluded by Section \ref{finalsec}, where discussion and conclusions are presented.

\section{Mathematical modeling of realistic sources}\label{framework}

We consider a laser that emits weak pulses characterized by coherent states $\ket{\alpha}$. The number of photons detected in a photon-counting measurement, $X$, is a random variable that follows a Poisson distribution. Thus, the probability of detecting exactly $n$ photons is given by
\begin{equation}
   \mathrm{Prob} (X = n)  = |\braket{n| \alpha }|^2= e^{- |\alpha|^2} \frac{ (|\alpha|^2)^n}{n!},
\end{equation}
where $|\alpha|^2$ denotes the expected value of the random variable $X$, i.e. $|\alpha|^2 = \mathbb{E} [X]$. For simplicity, we set $|\alpha|^2 \equiv \mathcal{N}$, which can be interpreted as the mean number of photons registered by the detector, meaning that $X \sim \mathrm{Pois} (\mathcal{N})$. Each photon is prepared in an identical quantum state, described by a density matrix $\rho_{x}$.

To reconstruct an unknown quantum state encoded in the photon's polarization degree of freedom, we implement a measurement scheme based on symmetric, informationally complete, positive operator-valued measures (SIC-POVMs) \cite{Renes2004,Fuchus2017}. For qubit tomography, the SIC-POVM involves four measurement operators that form a regular tetrahedron \cite{PaivaSanchez2010}. SIC-POVMs are known to be optimal for QST, providing a good balance between the number of measurements and the quality of state reconstruction.

Let us denote the measurement operators by $M_1, \dots, M_4$. This allows us to express the expected photon counts as:
\begin{equation}\label{eq1}
    \mathrm{exp}_{j} = \lceil \mathcal{N}\: 10^{- \gamma L} \:\tr M_j \rho_{x} \rfloor \hspace{0.5cm}\text{for}\hspace{0.5cm}j =1,\dots,4,
\end{equation}
where the symbol $\lceil a \rfloor$ denotes rounding $a$ to the nearest integer, since the number of photons cannot be fractional. The factor $10^{- \gamma L}$ represents the transmittance for a fiber characterized by length $L$ and an attenuation coefficient $\gamma$. The density matrix $\rho_{x}$ is unknown to the observer, and thus we parameterize it using the Cholesky decomposition, which depends on the dimension of the Hilbert space, cf. \cite{James2001,Altepeter2005}. For qubits, any density matrix is fully characterized by four real parameters $t_1, \dots, t_4$.

The formula \eqref{eq1} models photon counts according to the Born rule, the theoretical foundation of this measurement scheme. However, in practice, any measurement involves errors and uncertainty, meaning that the detected values may differ from theoretical expectations. In quantum optics, shot noise describes fluctuations in the number of photons reaching the detection system \cite{Hasinoff2014}. As a result, the measured counts, $\{m_j\}$, are statistically independent Poissonian random variables. To simulate an experimental scenario, we select photon numbers randomly from a Poisson distribution: $\widetilde{\mathcal{N}} \sim \mathrm{Pois} (\mathcal{N}\: 10^{- \gamma L})$. This procedure is repeated for each measurement operator, and the corresponding photon counts registered at the detector are given as:
\begin{equation}\label{eq2}
   m_j = \lceil \widetilde{\mathcal{N}}_j \: \tr M_j \rho_{in} \rfloor \hspace{0.5cm}\text{for}\hspace{0.5cm}j =1,\dots,4,
\end{equation}
where $\rho_{in}$ represents the input state of the photons. The value of $\widetilde{\mathcal{N}}_j$ is generated randomly, including the effects of fiber attenuation, since the number of photons in the coherent beam after propagation still follows a Poisson distribution. By assuming a specific form of $\rho_{in}$, we can generate noisy photon counts $m_j$ that simulate an experimental scenario.

In practice, beyond shot noise and attenuation, other phenomena contribute to errors in photon-counting schemes. As a specific example, we consider Raman scattering, which may occur due to the simultaneous transmission of quantum and classical light pulses. Based on Ref. \cite{Eraerds2010}, we compute the power of forward Raman scatter as:
\begin{equation}\label{Rampower}
    P_{for}^{Ram} = P_{in}\: L \: 10^{- \gamma L} \: d(\lambda) \: \Delta \lambda,
\end{equation}
where $P_{in}$ denotes the input classical power, $d(\lambda)$ is the effective Raman cross-section, and $\Delta \lambda$ is the spectral bandwidth. When considering Raman scattering, the spectral dependence of the Raman cross-section $d(\lambda)$ plays a significant role. The closer the quantum and classical channels are in wavelength, the more significant the Raman scattering noise. This dependence should be carefully accounted for, especially in systems where the wavelength separation between channels is minimal.

The formula in \eqref{Rampower} can be used to compute the number of scattered photons per second:
\begin{equation}\label{nRam}
    N^{Ram} = P_{for}^{Ram} \cdot \frac{\lambda_q}{h c },
\end{equation}
where $\lambda_q$ denotes the wavelength for the quantum channel. Assuming that each measurement requires establishing a temporal detection window $\tau$, the expected number of scattered photons is given by $\mathbb{E} [n^{R}] = \tau  N^{Ram}$. However, this is a stochastic process, and by using its expected value, we can simulate $n^{R}$, assuming it follows a Poisson distribution, i.e. $n^{R} \sim \mathrm{Pois} ( \tau  N^{Ram} )$.

Similarly, we can simulate noise photons that appear in the quantum channel due to crosstalk. The power of the crosstalk photons is given by \cite{Eraerds2010}
\begin{equation}
    P^{cr} =  P_{in}\: L \: 10^{- \gamma L} \: \xi, 
\end{equation}
where $\xi$ will be referred to as the crosstalk factor. The number of crosstalk photons per second, $N^{cr}$, can be computed analogously as in \eqref{nRam}. Then, for a detection window $\tau$, the number of crosstalk photons is a random variable such that $n^{cr} \sim \mathrm{Pois} ( \tau  N^{cr} )$. Finally, if two random variables follow a Poisson distribution, their sum is also a Poissonian random variable, with the expected value equal to the sum of the individual means. This allows us to consider the number of noise photons as a Poissonian variable, i.e., $n^{noise} \sim \mathrm{Pois} ( \tau (N^{Ram} + N^{cr}) )$.

Overall, the total number of photons arriving at the detection system, $N^{tot}$, can be simulated as:
\begin{equation}
    N^{tot} \sim \mathrm{Pois}  ( \mathcal{N}\: 10^{- \gamma L} + \tau  (N^{Ram} + N^{cr}) ).
\end{equation}

For the measurement in the polarization basis, we assume that for any measurement operator, $M_j$, statistically $50\%$ of scattered and crosstalk photons will pass through the measurement device. As a result, we obtain an expanded formula for the measured count:
\begin{equation}\label{measuredR}
   m_j = \lceil \widetilde{\mathcal{N}}_j \: \tr M_j \rho_{in} + 0.5\: (n_j^{R} + n_j^{cr}) \rfloor \hspace{0.5cm}\text{for}\hspace{0.5cm}j =1,\dots,4,
\end{equation}

The approach described above allows one to numerically generate photon counts corresponding to a realistic scenario for any input density matrix $\rho_{in}$. We then use the method of least squares (LS) to evaluate how well one can reconstruct the density matrix despite the uncertainties, cf. \cite{Acharya2019,Czerwinski2021}. This involves minimizing the function:
\begin{equation}\label{eq4}
    f_{LS} (t_1, t_2, t_3, t_4) = \sum_{j=1}^4 (\mathrm{exp}_j - m_j)^2,
\end{equation}
where $t_1, t_2, t_3, t_4$ denote the set of four real parameters that characterize the density matrix $\rho_x$. The measured count $m_j$ can be given by \eqref{eq2} or (\ref{measuredR}), depending on the noise scenario being analyzed.

To evaluate the performance of QST in the presence of fiber attenuation, we compute, for any input density matrix $\rho_{in}$, its fidelity with the estimated result $\rho_x$ \cite{Nielsen2000}:
\begin{equation}\label{eq5}
    F[\rho_{in}, \rho_{x}] := \left( \tr \sqrt{\sqrt{\rho_{in}} \rho_x\sqrt{\rho_{in}}} \right)^2,
\end{equation}
known as the Uhlmann-Jozsa fidelity, which measures the closeness of two quantum states \cite{Jozsa1994,Uhlmann1976}.

The efficiency of the framework should not be affected by the properties of input states $\rho_{in}$. Therefore, to find an indicator of the average performance, we first select a sample of input states, then apply the QST procedure to each $\rho_{in}$, and finally compute the average fidelity for the sample. Additionally, we calculate the sample standard deviation (SD) to quantify statistical dispersion. The sample SD is a natural choice for quantifying the uncertainty in QST affected by random noise since it effectively captures the variability in the performance of the framework. In our study, the fiber length is treated as an independent variable, implying that the average fidelity can be expressed as a function of $L$ and denoted by $F_{av} (L)$. Consequently, this figure of merit can be plotted to observe how the efficiency of QST changes as $L$ increases.

To ensure that the reconstructed density matrix $\rho_x$ remains physically valid, we employ the Cholesky decomposition. This approach is particularly advantageous as it guarantees that $\rho_x$ is positive semidefinite and has a unit trace, both of which are essential properties of any quantum state. The parameters derived from this decomposition correspond to physical characteristics of the quantum state, making them suitable for optimization in the least squares framework.

\section{Results and analysis}\label{resultsanalysis}

\subsection{Model Assumptions for Qubit Tomography}

For the numerical simulation, we select a sample of $200$ qubits parameterized as
\begin{equation}\label{qubit1}
\ket{\psi_{in}} =\begin{pmatrix}  \cos  \frac{\theta}{2} \\ \\ e^{ i \phi} \sin \frac{\theta}{2} \end{pmatrix},
\end{equation}
where $ 0\leq \phi < 2 \pi$ and $0\leq \theta \leq  \pi$. The sample is selected in such a way that the parameters $\phi$ and $\theta$ cover the full range. The simulations are limited to pure states since such states are predominantly encoded in the polarization degree of freedom and are widely used for quantum communications. For each sample qubit, we define an input density matrix $\rho_{in} = \ket{\psi_{in}} \bra{\psi_{in}}$ and generate noisy photon counts according to \eqref{eq2} or (\ref{measuredR}).

 \begin{table}[!h]
\def\arraystretch{2.0}
\begin{tabular}{|c|c|c|c|c|}
\hline
    \hspace{0.5cm}$\lambda_q$\hspace{0.5cm} 
      & \hspace{0.5cm}$\Delta \lambda$\hspace{0.5cm} &\hspace{0.5cm}$d(\lambda_q)$\hspace{0.5cm} & \hspace{0.75cm}$\gamma$\hspace{0.75cm} & \hspace{0.5cm}$\tau$\hspace{0.5cm} \\\hline
      $1548$ nm &  $45$ pm & \hspace{0.1cm}$1.5 \cdot 10^{-9} \:(\mathrm{km} \cdot \mathrm{nm})^{-1}$\hspace{0.1cm} & $0.2$ dB/km & $10^{-5}$ s\\\hline
\end{tabular}
 \caption{Parameters used in the QST simulations.}
\label{tab1}
\end{table}

Some parameters introduced in Section \ref{framework} are fixed and their values are provided in Table~\ref{tab1}. We assume that the carrier for the classical channel is $1550$ nm. The quantum channel is then adjusted to fall within the minimum value of the effective Raman cross-section. Based on Ref. \cite{Eraerds2010}, we select $\lambda_q = 1548$ nm with the corresponding value of the effective cross-section  $1.5 \cdot 10^{-9} (\mathrm{km} \cdot \mathrm{nm})^{-1}$. 

\subsection{QST Performance under Shot Noise and Attenuation}

\begin{figure}[!h] {\includegraphics[width=0.65\columnwidth]{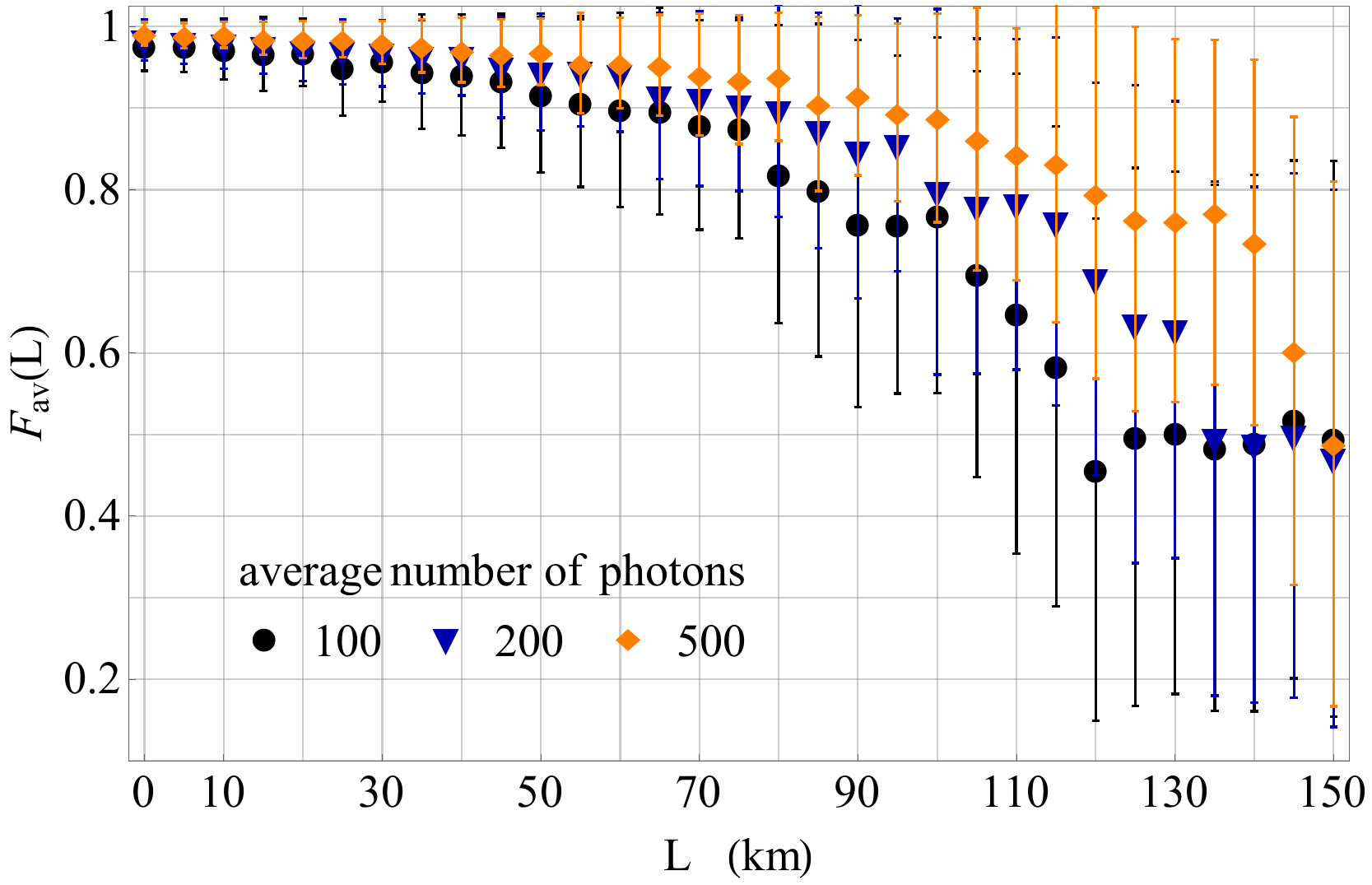}}
	\caption{Average fidelity of QST without Raman scattering and crosstalk.}
	\label{QST1}
\end{figure}

First, we test how attenuation and shot noise affect the efficiency of the framework. To do this, we generate noisy data according to \eqref{eq2} for the sample of input states $\rho_{in}$ and then implement the estimation technique.

Figure \ref{QST1} provides an insightful analysis of the efficiency of the framework under varying average photon numbers -- specifically, $\mathcal{N} = 100$, $200$, and $500$ photons -- while excluding Raman scattering and crosstalk effects. The results illustrate a clear relationship between the average photon number and the impact of attenuation on QST performance. The error bars in the plot represent the SD of the sample.

As the average photon number increases, the decline in QST efficiency due to attenuation becomes less pronounced. This trend demonstrates the increased robustness of the framework to attenuation when a higher photon number is employed. Conversely, for lower average photon numbers, the effect of attenuation is significantly more severe. This sensitivity to attenuation in lower photon regimes underscores a critical challenge in maintaining high-quality QST in scenarios where photon counts are limited.

The significant degradation in efficiency at lower photon numbers can be attributed to the increased relative impact of attenuation on the signal. With fewer photons, each photon carries a larger fraction of the total signal, making the measurement more susceptible to losses. As a result, the ability of the QST framework to accurately reconstruct quantum states diminishes more rapidly with attenuation when the initial photon number is low.

The results presented in Figure \ref{QST1} are quite analogous to the results presented in Ref.~\cite{CzerwinskiPRA}, where a similar problem was considered. However, the key difference lies in type of distribution used to simulate the attenuation effects -- in Ref.~\cite{CzerwinskiPRA} the binomial distribution was employed to model the number of photons that pass through the fiber.

\subsection{Impact of Raman Scattering and Crosstalk on QST Efficiency}

\begin{figure}[h]{\includegraphics[width=0.65\columnwidth]{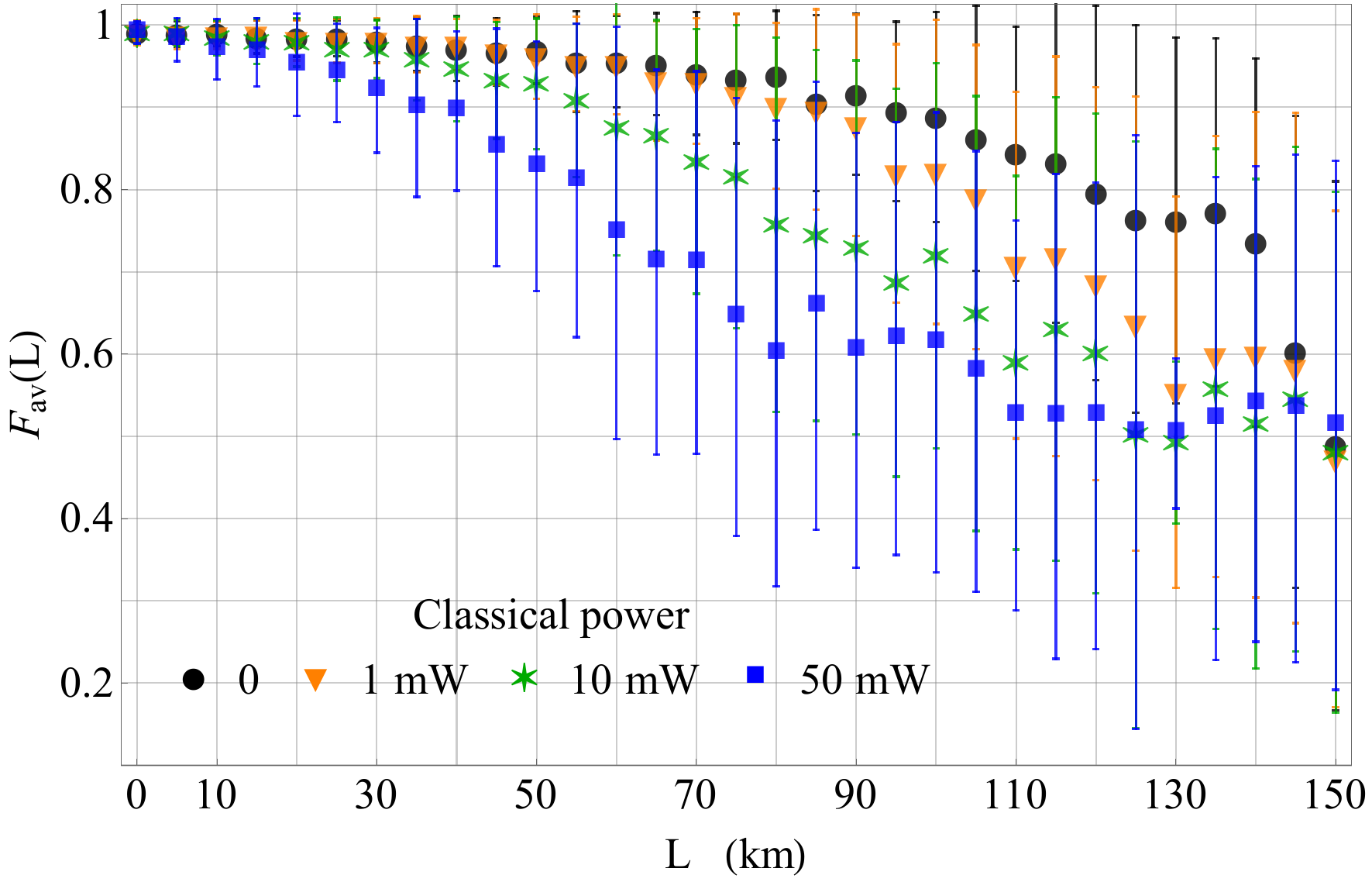}}
	\caption{Average fidelity of QST with Raman scattering. Zero crosstalk assumed, i.e. $\xi = 0.$}
	\label{QST2}
\end{figure}

Figure \ref{QST2} presents the efficiency of the QST framework, including the effects of Raman scattering but assuming zero crosstalk. The analysis involves four different levels of classical power $P_{in} = 0$, 1 mW, 10 mW, and 50 mW. As the classical power $P_{in}$ increases, there is a clear trend of more rapid degradation in the average fidelity $F_{av}(L)$. Specifically, with higher classical powers, the fidelity begins to drop off sooner as the fiber length increases.

For $P_{in} = 0$ (no classical power), the fidelity remains relatively high over a longer distance, indicating that without additional noise from the classical channel, QST is robust to fiber attenuation. As the classical power is increased to 1 mW, 10 mW, and eventually 50 mW, the onset of fidelity degradation occurs at progressively shorter fiber lengths. This confirms that a higher classical power induces more noise, particularly due to Raman scattering, which negatively impacts the quantum signal and therefore the accuracy of state estimation.

All the plots in Figure \ref{QST2}, regardless of the initial classical power, converge to a fidelity value of approximately $0.5$ as the fiber length increases. This means that for each case, we ultimately approach a situation when the framework results in random density matrices, for which the fidelity spans from $0$ to $1$ with the average value of $0.5$.

\begin{figure}[!h] {\includegraphics[width=0.7\columnwidth]{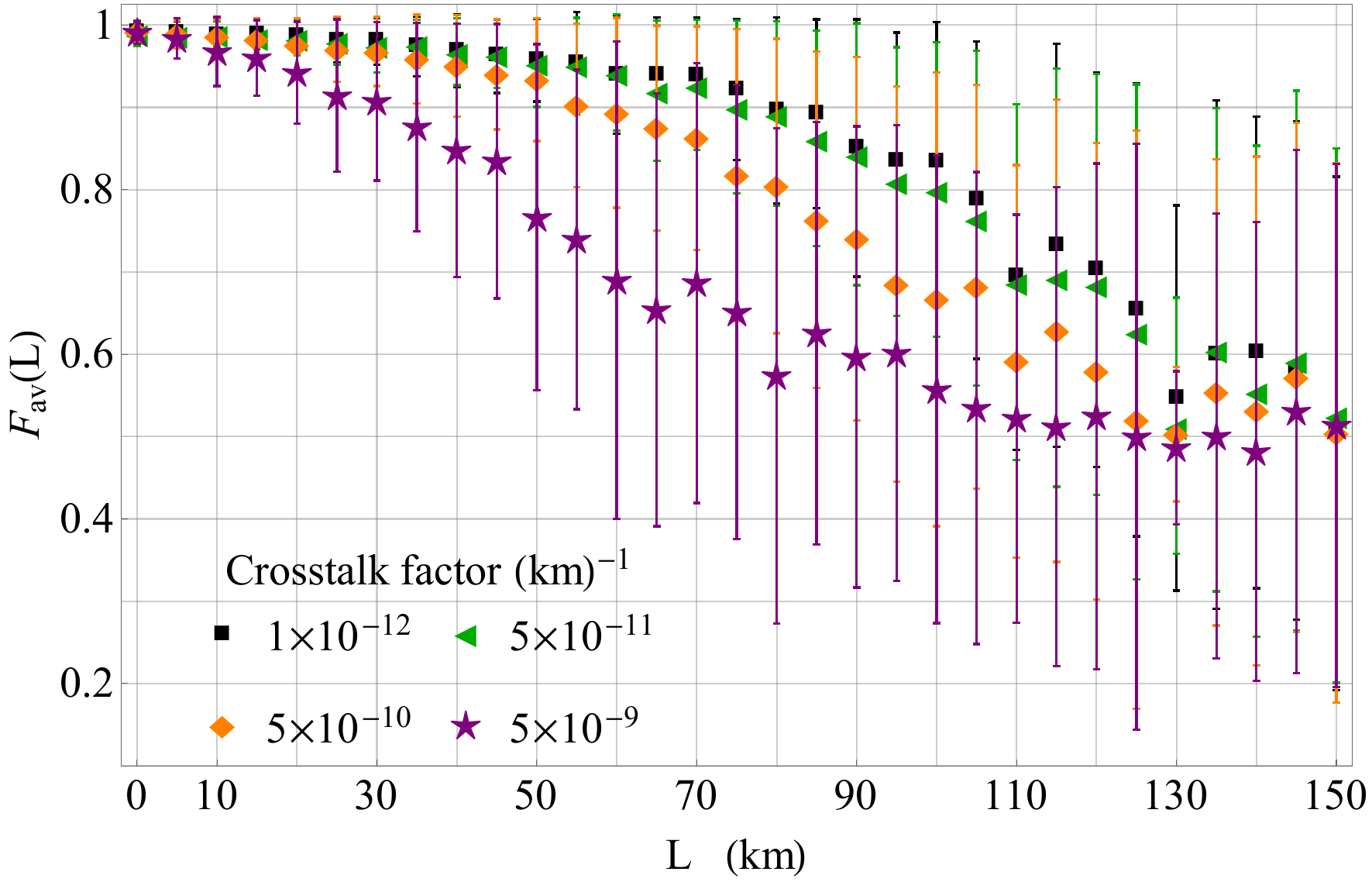}}
	\caption{Average fidelity of QST with Raman scattering and crosstalk. Classical power is fixed, $P_{in} = 1$ mW.}
	\label{QST3}
\end{figure}

Finally, Figure~\ref{QST3} illustrates the average fidelity QST as a function of fiber length $L$ for different values of the crosstalk factor, with the classical power fixed at $1$ mW. The analysis of these results reveals several important trends regarding the impact of crosstalk on the efficiency of QST.

For very low crosstalk factors, specifically $10^{-12}$ (km)$^{-1}$ and $5\cdot 10^{-11}$ (km)$^{-1}$, the impact of the crosstalk noise on QST efficiency is negligible. The plots for these values are nearly indistinguishable from the scenario where crosstalk is assumed to be zero, cf. Figure~\ref{QST2}. This suggests that at such low levels, the noise introduced by crosstalk is minimal and does not contribute significantly to the degradation of the quantum state reconstruction. In other words, QST maintains high fidelity across a wide range of fiber lengths, indicating that the framework is robust to these small levels of interference.

However, as the crosstalk factor increases to 
$5 \cdot 10^{-10}$ (km)$^{-1}$ and $5\cdot 10^{-9}$ (km)$^{-1}$, the effects on QST become much more pronounced. For these higher values of crosstalk, the average fidelity declines more rapidly as the fiber length increases. This rapid decline indicates that the noise generated by crosstalk becomes significant enough to interfere with the accurate reconstruction of the quantum state. The framework’s ability to perform reliable QST diminishes as the additional noise overwhelms the quantum signal, particularly over longer distances.

The results clearly demonstrate a threshold-like behavior where the crosstalk factor must exceed a certain value before it begins to significantly impact the fidelity of QST. Below this threshold, the QST process is relatively unaffected by crosstalk, but above it, the degradation is both rapid and severe. This suggests that in practical scenarios, managing crosstalk is critical. The findings highlight the importance of controlling crosstalk in quantum communication systems to ensure the reliable performance of QST, particularly in WDM environments where multiple signals are transmitted simultaneously.

\section{Discussion and Conclusions}\label{finalsec}

In this paper, we have developed and analyzed a numerical framework to simulate the efficiency of QST under realistic conditions, including shot noise, fiber attenuation, Raman scattering, and crosstalk. Our results demonstrate how these factors collectively influence the fidelity of quantum state reconstruction, with particular emphasis on the impact of classical power and crosstalk in WDM systems.

The transmittance factor $10^{-\gamma L}$ expresses the exponential dependence of photon loss on fiber length $L$. As the attenuation coefficient $\gamma$ varies with fiber material and wavelength, it becomes a critical determinant for the feasibility of long-distance quantum communication. Our findings suggest that as the fiber length increases, the accuracy of QST decreases more rapidly, especially when high classical powers are transmitted simultaneously. Future research could extend this work by exploring different types of fibers, where variations in the attenuation coefficient and the effective Raman cross-section may lead to distinct effects on the efficiency of QST frameworks. Such studies could provide valuable insights for optimizing quantum communication over a range of fiber types and operational conditions.

Furthermore, while our primary focus has been on the effects of shot noise, Raman scattering, and crosstalk, it is crucial to consider additional sources of error in photon-counting detectors, such as detector inefficiencies and dark counts. These factors, although not modeled in the current framework, can introduce further noise and uncertainty, complicating the QST process. Incorporating these sources of error into future frameworks could lead to even more complex frameworks, making them more applicable to real-world scenarios where such imperfections are inevitable.

In addition to the specific noise factors studied, our results highlight the importance of carefully managing crosstalk in WDM systems. The threshold-like behavior observed in our simulations indicates that crosstalk must be kept below certain levels to prevent significant degradation in QST performance. This insight could inform the design of quantum communication networks, where minimizing crosstalk is essential for maintaining high fidelity in quantum state estimation.

Overall, this study provides a comprehensive understanding of the challenges posed by realistic noise sources in QST and offers a foundation for further research aimed at improving the reliability and accuracy of QST in practical quantum communication systems. Future work could explore advanced error mitigation techniques, alternative quantum state reconstruction methods, and the integration of these frameworks into experimental setups to validate and refine the theoretical predictions presented here. By addressing these challenges, we can move closer to realizing robust quantum communication networks capable of operating effectively under real-world conditions.

In addition, future research on QST could be influenced by recent findings on generalized measurements, which have demonstrated the potential for improved randomness certification and enhanced state estimation techniques \cite{Mironowicz2024}. Moreover, quantum states encoded in a photon's degrees of freedom other than polarization could be considered. For instance, encoding quantum information into the frequency of photons has recently been demonstrated as an effective approach to transmitting quantum information in the presence of classical signals \cite{Philip2024}. Finally, the method introduced in the present paper can be extended to investigate the efficiency of other state characterization techniques, such as quantum state discrimination \cite{Khan2024}.


\begin{thebibliography}{1}

\bibitem{Paris2004}
M. Paris and J. \v{R}eh\'{a}\v{c}ek (eds.), \textit{Quantum State Estimation} (Springer, Heidelberg, 2004).

\bibitem{Toninelli2019}
E. Toninelli, B. Ndagano, A. Valles, B. Sephton, I. Nape, A. Ambrosio, F. Capasso, M.J. Padgett, and A. Forbes, Concepts in quantum state tomography and classical implementation with intense light: A tutorial. Adv. Opt. Photon. {\bf 11}, 67–134 (2019).

\bibitem{Czerwinski2022}
A. Czerwinski, Selected Concepts of Quantum State Tomography. Optics {\bf 3}, 268-286 (2022).

\bibitem{Gebhart12023}
V. Gebhart, R. Santagati, A.A. Gentile, E.M. Gauger, D. Craig, N. Ares, L. Banchi, F. Marquardt, L. Pezzè, and C. Bonato, Learning quantum systems. Nat Rev Phys {\bf 5}, 141-156 (2023).

\bibitem{Walther2005}
P. Walther, K.J. Resch, T. Rudolph, E. Schenck, H. Weinfurter, V. Vedral, M. Aspelmeyer, and A. Zeilinger, Experimental one-way quantum computing. Nature {\bf 434}, 169–176 (2005).

\bibitem{Kok2007}
P. Kok, W.J. Munro, K. Nemoto, T.C. Ralph, J.P. Dowling, and G.J. Milburn, Linear optical quantum computing with photonic qubits. Rev. Mod. Phys. {\bf 79}, 135–174 (2007).

\bibitem{Wei2022}
M. Wei, C.-H. Zhang, J. Li, J.-L. Zhu, and Q. Wang, Experimental demonstration of tomography-based quantum key distribution. Opt. Lett. {\bf 47}, 6285–6288 (2022).


\bibitem{Slaoui2023}
A. Slaoui, N. Ikken, L. Btissam Drissi, and R. Ahl Laamara, {\it Quantum Communication Protocols: From Theory to Implementation in the Quantum Computer}, [in:] {\it Quantum Computing - Innovations and Applications in Modern Research}, edited by B. Carpentieri, (IntechOpen, London, UK, 2023).

\bibitem{Slaoui2024}
A. Slaoui, M. El Kirdi, R. Ahl Laamara, M. Alabdulhafith, S.A. Chelloug, and A.A. Abd El-Latif, Cyclic quantum teleportation of two-qubit entangled states by using six-qubit cluster state and six-qubit entangled state. Sci Rep {\bf 14}, 15856 (2024).

\bibitem{Ali2024}
A. Ali, S. Al-Kuwari, and S. Haddadi, Trade-off relations of quantum resource theory in Heisenberg models. Phys. Scr. {\bf 99}, 055111 (2024).

\bibitem{James2001}
D. F. V. James, P. G. Kwiat, W. J. Munro, and A. G. White, Measurement of qubits. Phys. Rev. A \textbf{64}, 052312 (2001).

\bibitem{Altepeter2005}
J. B. Altepeter, E. R. Jeffrey, and P. G. Kwiat, Photonic State Tomography. Adv. At. Mol. Opt. Phys. \textbf{52}, 105-159 (2005).

\bibitem{Horn2013}
R. T. Horn, P. Kolenderski, D. Kang, P. Abolghasem, C. Scarcella, A. D. Frera, A. Tosi, L. G. Helt, S. V. Zhukovsky, J. E. Sipe, G. Weihs, A. S. Helmy, and T. Jennewein, Inherent polarization entanglement generated from a monolithic semiconductor chip. Sci. Rep. \textbf{3}, 2314 (2013).

\bibitem{Temporao2024}
G.P. Temporão, P. Ripper, T.B. Guerreiro, and G.C. do Amaral, Two-photon quantum state tomography of photonic qubits. Phys. Rev. A {\bf 109}, 022402 (2024).

\bibitem{Hasinoff2014}
S. W. Hasinoff, {\it Photon, Poisson noise}, [in:] \textit{Computer Vision}, edited by K. Ikeuchi, (Springer, Boston, MA, 2014), pp. 608-610.

\bibitem{Eraerds2010}
P Eraerds, N Walenta, M Legré, N Gisin, and H Zbinden, Quantum key distribution and 1 Gbps data encryption over a single fibre. New J. Phys. \textbf{12}, 063027 (2010).

\bibitem{Frohlich2015}
B. Fr\"ohlich, J.F. Dynes, M. Lucamarini, A.W. Sharpe, S.W.B. Tam, Z. Yuan, and A.J. Shields, Quantum secured gigabit optical access networks. Scientific reports {\bf 5}, 18121 (2015).

\bibitem{CzerwinskiQIC}
A. Czerwinski, Entanglement characterization by single-photon counting with random noise.  Quantum Inf. Comput. {\bf 22}, 1-16 (2022).

\bibitem{Wang2022}
H. Wang, K. He, Y. Hao, and S. Yang, Quantum tomography with Gaussian noise. Quantum Inf. Comput. {\bf 22}, 1144-1157 (2022).

\bibitem{CzerwinskiPRA}
A. Czerwinski and J. Szlachetka, Efficiency of photonic state tomography affected by fiber attenuation. Phys. Rev. A {\bf 105}, 062437 (2022).

\bibitem{Pedram2024}
A. Pedram, V.R. Besaga, F. Setzpfandt, Ö.E. Müstecaplıoğlu, Nonlocality enhanced precision in quantum polarimetry via entangled photons. Adv Quantum Technol., 2400059 (2024) [Online Version of Record before inclusion in an issue] \doi{10.1002/qute.202400059}

\bibitem{Mukherjee2006}
B. Mukherjee, {\it Optical WDM networks}, (Springer Science \& Business Media, New York, NY, 2006).

\bibitem{Gobby2004}
C. Gobby, Z.L. Yuan, and A.J. Shields, Quantum key distribution over 122 km of standard telecom fiber. Appl. Phys. Lett. {\bf 84}, 3762–3764 (2004)

\bibitem{Tomita2010}
A. Tomita, K.-I. Yoshino, Y. Nambu, A. Tajima, A. Tanaka, S. Takahashi, W. Maeda, S. Miki, Z. Wang, M. Fujiwara, and M. Sasaki, High speed quantum key distribution system. Opt. Fiber Technol. {\bf 16}, 55-62 (2010).

\bibitem{Cao2019}
Y. Cao, Y. Zhao, J. Wang, X. Yu, Z. Ma, and J. Zhang, Cost-Efficient Quantum Key Distribution (QKD) Over WDM Networks.  J. Opt. Commun. Netw. {\bf 11}, 285-298 (2019)


\bibitem{Zhou1996}
J. Zhou, R. Cadeddu, E. Casaccia, C. Cavazzoni, and M.J. O'Mahony, Crosstalk in multiwavelength optical cross-connect networks. J. Light. Technol {\bf 14}, 1423–1435 (1996).

\bibitem{Monroy2013}
I.T. Monroy, and E. Tangdiongga, {\it Crosstalk in WDM communication networks}, (Springer Science \& Business Media, New York, NY, 2013).

\bibitem{Renes2004}
J. M. Renes, R. Blume-Kohout, A. J. Scott, and C. M. Caves, Symmetric informationally complete quantum measurements. J. Math. Phys. \textbf{45}, 2171–2180 (2004).

\bibitem{Fuchus2017}
Ch. A. Fuchs, M. C. Hoang, and B. C. Stacey, The SIC Question: History and State of Play. Axioms \textbf{6}, 21 (2017).

\bibitem{PaivaSanchez2010}
C. Paiva-S\'{a}nchez, E. Burgos-Inostroza, O. Jim\'{e}nez, and A. Delgado, Quantum tomography via equidistant states. Phys. Rev. A \textbf{82}, 032115 (2010).

\bibitem{Acharya2019}
A. Acharya, T. Kypraios, and M. Guta, A comparative study of estimation methods in quantum tomography. J. Phys. A: Math. Theor. \textbf{52}, 234001 (2019).

\bibitem{Czerwinski2021}
A. Czerwinski, K. Sedziak-Kacprowicz, and P. Kolenderski, Phase estimation of time-bin qudits by time-resolved single-photon counting. Phys. Rev. A \textbf{103}, 042402 (2021).

\bibitem{Nielsen2000}
M.~A.~Nielsen and I.~L.~Chuang, \textit{Quantum Computation and Quantum Information}, (Cambridge University Press, Cambridge, 2000).

\bibitem{Jozsa1994}
R. Jozsa, Fidelity for Mixed Quantum States. J. Mod. Opt. \textbf{41}, 2315-2323 (1994).

\bibitem{Uhlmann1976}
A. Uhlmann, The “transition probability” in the state space of a $\star$-algebra. Rep. Math. Phys. \textbf{9}, 273-279 (1976).

\bibitem{Mironowicz2024}
P. Mironowicz, M. Grünfeld, and M. Bourennane, Generalized measurements on qubits in quantum randomness certification and expansion. Phys. Rev. Applied {\bf 22}, 044041 (2024).

\bibitem{Philip2024}
P. Rübeling, J. Heine, R. Johanning, and M. Kues, Quantum and coherent signal transmission on a single-frequency channel via the electro-optic serrodyne technique. Sci. Adv. {\bf 10}, eadn8907 (2024)

\bibitem{Khan2024}
A.K. Khan, Y.H. Dar, E.C. Vagenas, S.S. Wani, S. Al-Kuwari, and M. Faizal, Effects of underlying topology on quantum state discrimination. Eur. Phys. J. C {\bf 84}, 240 (2024). 

\end{thebibliography}
\end{document}